\documentclass[10pt]{article}

\usepackage{amsmath}
\usepackage{array}
\usepackage{appendix}
\usepackage{graphicx}
\usepackage{amsfonts}
\usepackage{amssymb}
\usepackage{mathrsfs}
\usepackage{yfonts}
\usepackage{euscript}
\usepackage{upgreek}
\usepackage{slantsc}
\usepackage{calligra}
\usepackage[T1]{fontenc}
\usepackage{epsf}
\usepackage{latexsym}

\usepackage{tipa}

\def\be{\begin{equation}}
\def\ee{\end{equation}}
\def\beq{\begin{equation}}
\def\eeq{\end{equation}}
\def\bea{\begin{eqnarray}}
\def\eea{\end{eqnarray}}

\def\bu{\noindent$\bullet$}

\def\hat{\widehat}

\def\!{\hspace{-0.3em}}

\def\mP{\mbox{P}}

\def\mS{\mbox{S}}

\def\mk{\mbox{k}}

\def\bigmu{\mbox{\Large $\mu$}}

\def\fG{\mbox{\sffamily G}}

\def\fQ{\mbox{\sffamily Q}}
\def\fR{\mbox{\sffamily R}}
\def\fS{\mbox{\sffamily S}}

\def\bn{\mbox{\bf n}}

\def\bM{\mbox{\bf M}}
\def\bl{\mbox{\bf l}}

\def\bM{\mbox{{\bf M}}}
\def\bM{\mbox{{\bf M}}}

\def\bn{\mbox{{\bf n}}}

\def\sa{\mbox{\scriptsize a}}

\def\scc{\mbox{\scriptsize c}}

\def\se{\mbox{\scriptsize e}}

\def\si{\mbox{\scriptsize i}}

\def\sll{\mbox{\scriptsize l}}  
\def\sm{\mbox{\scriptsize m}}
\def\sn{\mbox{\scriptsize n}} 
 
\def\sp{\mbox{\scriptsize p}}

\def\sr{\mbox{\scriptsize r}}
\def\sss{\mbox{\scriptsize s}}  
\def\st{\mbox{\scriptsize t}}

\def\sB{\mbox{\scriptsize B}}

\def\sJ{\mbox{\scriptsize J}}
\def\sK{\mbox{\scriptsize K}}

\def\sO{\mbox{\scriptsize O}}
\def\sP{\mbox{\scriptsize P}}

\def\sS{\mbox{\scriptsize S}}

\def\sU{\mbox{\scriptsize U}}

\def\sW{\mbox{\scriptsize W}}

\def\sfa{\mbox{\sffamily{\scriptsize a}}}

\def\sfQ{\mbox{\sffamily{\scriptsize Q}}}
\def\sfR{\mbox{\sffamily{\scriptsize R}}}

\def\sbM{\mbox{{\bf \scriptsize M}}}

\def\lt{\mbox{\Large $t$}}

\def\pa{\partial}
\def\d{\textrm{d}}

\def\Last{\mbox{\Large$\ast$}}                
\def\5Star{\mbox{\Large$\star$}}              
\def\Rec{\mbox{\textcircled{$\star$}}}        

\def\Spade{\mbox{$\spadesuit$}}           
\def\Club{\mbox{$\clubsuit$}}             

\def\cr{\mbox{\scriptsize{\bf $\mbox{ } \times \mbox{ }$}}}

\def\sumi3{\sum\mbox{}_{\mbox{}_{\mbox{\scriptsize $i$=1}}}^3}

\def\sumj3{\sum\mbox{}_{\mbox{}_{\mbox{\scriptsize $j$=1}}}^3}
\def\sumk3{\sum\mbox{}_{\mbox{}_{\mbox{\scriptsize $k$=1}}}^3}

\def\half{\mbox{$\frac{1}{2}$}}

\begin{document}

\begin{titlepage}

\vspace{.7in}
\begin{center}
\large{\bf QUANTUM COSMOLOGY WILL NEED TO BECOME A NUMERICAL SUBJECT}\normalsize 

\vspace{.15in}

{\large \bf Edward Anderson}\footnote{ea212@cam.ac.uk}

\vspace{.15in}

\large {\em DAMTP, Cambridge}. \normalsize

\begin{abstract}

The inhomogeneous fluctuations that underlie structure formation -- galaxies and CMB hot spots -- might have been seeded by quantum cosmological fluctuations, 
as magnified by some inflationary mechanism.  
The Halliwell--Hawking model for these, as a lower-energy semiclassical limit, is expected to be shared by many theories.  
E.g. an $O((H/m_{\sp\sll})^2)$ suppression of power at large scales results from this.
This model contains/suppresses very many terms; we want a qualitative understanding of the meaning of these terms and of different regimes resulting from different combinations of them.  
I study this with toy models that have tractable mathematics: minisuperspace and, especially, relational particle mechanics. 
In the present Seminar, I consider in particular averaged terms with some lessons from Hartree--Fock approach to Atomic and Molecular Physics.
One needs to anchor this on variational principles; treating the subsequent equations is a {\sl numerical} venture.

\end{abstract}

\end{center}

\noindent Invited Seminar at 'XXIX-th International Workshop on High Energy Physics: New Results and Actual Problems in Particle \& Astroparticle Physics and Cosmology' (Moscow 2013).

\section{Introduction}\label{Intro}

\noindent The inhomogeneous fluctuations that underlie structure formation -- galaxies and CMB hot spots -- might have been seeded by quantum cosmological fluctuations 
\cite{HallHaw},\footnote{See Appendix A for an outline of Halliwell and Hawking's (HH) approach to this.} 
as magnified by some inflationary mechanism \cite{inflation-13}.  
While different fundamental theories might give different quantum cosmological fluctuations, these would mostly be expected to be similar in the semiclassical regime through 
this being insensitive to more high-energy details such as spacetime discreteness, noncommutativity or branes.
Kiefer et al \cite{Kiefer-Recent} have shown that is a $O((H/m_P)^2)$ effect (for Hubble parameter $H$ and Planck mass $m_{\sP\sll}$, so this is a 1 part in $10^{10}$ effect) 
that enters as a suppression of power at large scales for plain Wheeler--DeWitt geometrodynamics.

The HH scheme is, admittedly, not ideal from the perspective of modern cosmology (which goes as far as second-order perturbations albeit at the classical level). 
But it is a start and it has some complementary values.   
For instance it is a semiclassical emergent time resolution of the Frozen Formalism Problem facet of the Problem of Time \cite{K92I93APOTAPOT2, FileR}. 
(Moreover, it is a Machian resolution \cite{ARel2, ACos2}: {\sl time is abstracted from change} and, furthermore, {\sl all changes are given an opportunity to contribute}.)
Also, it is a more fully quantum scheme than is usually used in modelling inflation, including in the sense of how {\sl whole-universe} QM is to be {\sl interpreted}.

We have to confront the issue that schemes like HH's exclude a lot of the semiclassical approach's terms.
Is this justified, and why in detail? 
Surely sometimes one finds one has to keep some terms in certain regimes, this is not unlike handling any other many-term equation in physics. 
Going in another direction (multiple theories' semiclassical schemes), can (perhaps higher order term contributions in) semiclassical quantum cosmology discern between different theories?

Here it helps to consider first toy models of the HH system, such as minisuperspace \cite{BI75, AMSS1, AMSS2} or relational particle mechanics (RPM) \cite{FileR, ACos2}.  
RPM's \cite{BB82, AF+tri, QuadI, AK13, FileR} have two {\sl midi}superspace like features -- nontrivial linear constraints and a nontrivial notion of structure/inhomogeneity: 
clumping, or, more generally, {\it shape}.   
Here small-inhomogeneity GR modelled by small changes in the shapes made by the point-particles, whilst the scale dynamics of the model can be attuned to match conventional Cosmology's.  
(This is in close parallel \cite{FileR} to how \cite{BGS} achieve this from Newtonian Cosmology.)
RPM's are also also a stepping-stone for importing Molecular Physics concepts and techniques into Quantum Cosmology \cite{FileR}.  
This is due to their classical-kinematical (if not quantum) parallels with molecules {\sl in a small corner of} the universe, for all that RPM's are now {\sl whole-universe} models.  
Their main previously-known limitations are in not having the following.
1) An indefinite kinetic term like GR's. 
2) A meaningful equation of state. 
3) A manifestation of microlensing 
(Their principal value is as qualitative toy models of {\sl GR in dynamical form} rather than as detailedly accurate cosmologies.)  
I am currently investigating insights from both minisuperspace and RPM's for HH type models of actual Quantum Cosmology.

An outline of the rest of this Seminar is as follows.
In Sec 2, I outline the classical and quantum scaled triangleland RPM.
I then derive the corresponding semiclassical approach equations in Sec 3, and give a term-by-term analysis of these in Sec 4. 
Appendix A compares this in outline with the HH system itself.
In Sec 5, I concentrate on averaged correction terms. 
I demonstrate that these are not always small.  
I make a parallel with the Hartree--Fock scheme from Atomic and Molecular Physics.  
Here, averages could not be neglected if one was to reproduce anything like the accurate spectra.  
Moreover, their inclusion drove one from analytic to numerical computation. 
This is a so-called {\it self-consistent scheme} of heavily coupled integro-differential equations.
The treatment of such schemes being anchored on variational principles, I provide one for the semiclassical approach's light subsystem's wave equation in Sec 6. 
I briefly comment on providing a separate variational principle for the heavy subsystem, and the further difficulties in finding a variational principle for the coupled system.
I also outline how Hartree--Fock type theory remains useable and known for time-dependent and field-theoretic applications as are to be expected in the study of 
HH-like semiclassical quantum cosmology.  
I conclude in Sec 7.

\end{titlepage}

\section{Quantum relational triangle toy model}\label{Toy}

\noindent 3 particles in 2-$d$ (or higher) have 
shape space $\fS(3, 2)$ = $\mathbb{CP}^1 = \mathbb{S}^2$          (topologically and metrically) and 
relational space -- including scale -- $\fR(3, 2) = \mathbb{R}^3$ (topologically but metrically only up to a conformal factor).  
See Fig 1 for how to sequentially free oneself of absoluteness from one's coordinatization; all of this can be done in terms of reduction at the Lagrangian level too.  
This geometrical simpleness greatly enhances solvability of many analogues of Problem of Time approaches, semiclassical quantum cosmology correction terms and other foundational issues 
that are nontrivially present for this model \cite{FileR}.  
%
{            \begin{figure}[ht]
\centering
\includegraphics[width=0.6\textwidth]{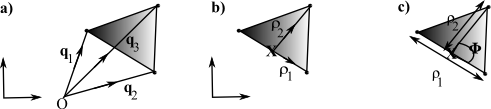}
\caption[Text der im Bilderverzeichnis auftaucht]{        \footnotesize{Progression of coordinate systems for the triangle. 
a) are particle position coordinates relative to an absolute origin and absolute axes.  
b) are relative Jacobi interparticle cluster separations; X denotes the centre of mass of particles 1 and 2; note that these coordinates are still relative to absolute axes.
Then the configuration space radius ('hyperradius' in molecular physics literature), $\rho := \sqrt{\rho_1^2 + \rho^2_2}$.    
Also, $n_i := \rho_i/\rho$. 
c) are scaled relational coordinates (ie no longer with respect to any absolute axes either).  
Pure-shape coordinates are then the relative angle $\Phi$ and some function of the ratio $\rho_2/\rho_1$; in particular, $\Theta := 2\mbox{arctan}(\rho_2/\rho_1)$.} }
\label{RPM-coordi} \end{figure}          }

Next, kinematical quantization for this model \cite{AK13} is a subcase of Isham's prescription \cite{I84} for configuration spaces $\fQ$ of group-quotient form $\fG_1/\fG_2$.
This requires the semidirect product $\mbox{\textcircled{S}}$ of the corresponding canonical group $\fG_{\scc\sa\sn}$ (which is $\fQ$'s isometry group in this subcase)
and the dual of a linear space within which the group orbits are realized, $V^*$ (which is just $V$ for finite examples such as the present one).  
Thus for triangleland one has \{$SU(2) \mbox{\textcircled{S}} \mathbb{R}^3\} \mbox{\textcircled{S}} \mathbb{R}^3$, 
the linear space's basis of 3 objects obeying the on-sphere condition $X^2 + Y^2 + Z^2 = 1$.
$X$, $Y$ and $Z$ are, moreover, of somewhat non-obvious form in this case, being `Dragt coordinates': $2\bn_1\cdot\bn_2$,  $2\{\bn_1\cr\bn_2\}_3$ and $n_2^2 - n_1^2$.
These are related to the Hopf map, or, most succinctly \cite{QuadI}, quadratic forms built out of the Pauli matrices via the well-known $SO(3)$ vector to $SU(2)$ matrix relation. 
Hence this `$\mathbb{R}^3$' is better envisaged as the spaces of irreducible homogeneous polynomials (IHP) of degree 2 over the complex numbers.

Finally, the triangleland quantum wave equation, which we choose to be conformal operator-ordered \cite{Magic} for further relational reasons \cite{Banal, FileR}, is \cite{AK13} 
\beq
-\hbar^2\{\pa^2_{\rho} + 2\rho^{-1}\pa_{\rho} + \rho^{-2}\{\triangle_{\mathbb{CP}^1} - 3/2\}\}\Psi = 2\{E_{\sU\sn\si} - V(\rho, \Theta, \Phi)\}\Psi
\label{Gilthoniel}
\eeq
Here $E_{\sU\sn\si}$ is the energy of the model universe, taken to be fixed, and $V$ is the potential function.  
This equation scale--shape separates, with the scale part solved for the general free and isotropic-HO potential cases \cite{AK13}, 
giving solutions in terms of Bessel functions and associated Laguerre polynomials respectively.  
The shape part gives spherical harmonics  $Y_{\sS\sss}(\Theta, \Phi)$. 
(Here, the quantum numbers indicate shape momenta: mixed relative angular momentum and relative dilational momentum, 
though in a few bases have s is a pure relative angular momentum and thus denoted by j \cite{AF+tri, FileR}.)

\section{Semiclassical scheme for this model}

\noindent{\bf Born--Oppenheimer (BO) scheme and its quantum-cosmological analogue} 
I take this to mean the ansatz $\psi(h,l) = \phi(h)|\chi(h,l)\rangle$ alongside the following approximations.  
\noindent 1) For $\widehat{C} := \hat{H}  -  \hat{T}_{h}$: the complement of the heavy kinetic term, the $|\chi\rangle$-wavefunction expectation value 
(`$l$-averaged': integrated over the $l$ degrees of freedom's configuration space, $\fQ_l$) is 
\beq
c := \langle\chi|\widehat{C}|\chi\rangle = \int_{\sfQ_l}
     \chi^{*}(h, l^{\sfa}) \, \widehat{C}(h, l^{\sfa}, p^{l}_{\sfa}) \, \chi(h, l^{\sfa})\, \mathbb{D} \fQ_l \mbox{ } .  
\eeq 
The associated integration here is over the $l$ degrees of freedom, e.g. in the present context, the shape space, with measure $\mathbb{D}l = \mathbb{D}\mS$, 
for $\mS^{\sfa}$ the shape degrees of freedom.  
Then using $c_{\alpha\alpha}$ as a more complete name for the above $c$ (where $\alpha$ is a multi-index running over the quantum numbers of $|\chi\rangle$), 
there is an obvious off-diagonal equivalent 
\be
c_{\alpha\beta} := \langle\chi_{\alpha}|\hat{C}|\chi_{\beta}\rangle \mbox{ } .  
\ee 
Then the {\it BO approximation} alias `diagonal dominance condition' is that 
\be
\mbox{for } \alpha \neq \mk \mbox{ } , \left|c_{{\alpha}{\beta}}/c_{{\alpha}{\alpha}}  \right| =: \epsilon_{\sB\sO} << 1 \mbox{ } .  
\ee
Henceforth, $\langle O \rangle$ denotes $\langle\chi|O|\chi\rangle$. 

\noindent 2) {\sl Adiabatic approximations}.
Because the ansatz is a product in which both factors depend on $h$, the $h$-derivatives acting upon it produce multiple terms by the product rule.  
In particular, both BO's own atomic-and-molecular-physics case and the quantum-cosmological case contain second derivatives in $h$. 
Then the product rule produces three terms, schematically 
\beq
|\chi\rangle \pa_{h}^2\psi \mbox{ } , \mbox{ } \mbox{ }  \pa_{h}\psi \pa_{h}  |\chi\rangle \mbox{ } , \mbox{ } \mbox{ } \psi\pa^2_{h}  |\chi\rangle \mbox{ } .  
\eeq
The first is always kept.  
BO themselves discarded the next two as being far smaller than the first (this is due to a first kind of {\sl adiabatic approximation}. 
By this, $h$-changes in $\chi$ are considered to be much smaller than $h$-changes in $\psi$).   
However, as explained below, the emergent semiclassical time approach to the Problem of Time requires at least one such cross-term to be kept. 
If there is a linear derivative term in the ${\cal H}\Psi = 0$ (e.g. from curvilinear coordinates or curved space), the product rule likewise produces two further terms,
\beq
|\chi\rangle  \pa_{h}\psi \mbox{ } , \mbox{ }  \mbox{ } \psi\pa_{h}  |\chi\rangle \mbox{ } .
\eeq  
The second of these then discarded due to the above kind of adiabatic assumption.

In addition to classical adiabatic terms of form $\epsilon_a = \omega_h/\omega_l$, there are two further `pure forms' that adiabaticity can take at the quantum level.  
\noindent a($l$): quantities that are small via $|\chi\rangle$ being far less sensitive to changes in $h$-subsystem physics than to changes in $l$-subsystem physics.  
\noindent a($m$): quantities that are small via $|\chi\rangle$ being far less sensitive to changes in $l$-subsystem physics than $\psi$ is sensitive to changes in 
$h$-subsystem physics.

\noindent Note 1) The $l$ stands for `internal to the $l$-subsystem' and the $m$ for `mutual between the $h$ and $l$ subsystems'.)  

\noindent Note 2) Massar and Parentani's work on inclusion of non-adiabatic effects \cite{MP98} in the minisuperspace arena is similar in spirit to the present Sec 
as regards considering the qualitative effects of retaining terms usually neglected in the Quantum Cosmology literature.  
This results in 1) couplings between expanding and contracting universe states.
2) A quantum-cosmological realization of the Klein paradox (backward-travelling waves generated from an initially forward-travelling wave). 

\mbox{ } 

\noindent{\bf The WKB scheme}.  
This to consist of the subsequent ansatz $\phi(h) = \mbox{exp}(iS(h)/\hbar)$ alongside the following approximations.

\noindent Rewrite the principal function $S$ by isolating a heavy mass $M$, $S(h) = M F(h)$,  
%
%
the {\it WKB approximation} is the negligibility of second derivatives, 
\be
\left|\frac{\hbar}{M}\frac{\pa_{h}^2F}{|\pa_{h}F|^2}\right| << 1 \mbox{ } \mbox{ with associated dimensional analysis expression } \mbox{ } \mbox{ } 
\hbar/MF =: \epsilon_{\sW\sK\sB^{\prime}} << 1 \mbox{ } . 
\eeq
Applying the well-known reinterpretation of $S$ as classical action, the above condition is to be interpreted as (quantum of action) $<<$ (classical action), 
which has clear semiclassical connotations.  

\mbox{ } 

\noindent The Semiclassical Approach to Quantum Cosmology's WKB ansatz is contentious \cite{Zeh, K92I93APOTAPOT2}.
In the present program, I deal with it in a combined Machian semiclassical histories records way in \cite{AHall, FileR, CapeTown12} 
based on the earlier non-Machianly interpreted work of Halliwell \cite{H03}.

\mbox{ } 

\noindent{\bf The BO--WKB scheme's scale--shape split RPM $h$- and $l$-equations}. 
\noindent I choose to use the reduced, scale $\rho$ = $h$, shape $(\Theta, \Phi)$ = $l^{\sfa}$ triangleland for further work in detail in this Seminar. 
Substitute the above brace of ans\"{a}tze into (1) and apply $\langle \chi | \times$ to form the {\it h-equation} \cite{AK13} 
$$ 
0 =     \{\pa_{h}S\}^2 - i\hbar \, \pa_{h}\mbox{}^2S - 2i\hbar \{\langle \pa_{h} \rangle + h^{-1} \} \pa_{h}S 
    - \hbar^2\big\{  \langle   \pa_{h}\mbox{}^2  \rangle + 2h^{-1}  \langle \pa_{h} \rangle + h^{-2}\{\langle \triangle_{l} \rangle - 3/2\}\}   
 + 2\{V_{h}(h) + \langle  V_{l}(l) + J(h, l^{\sfa}) \rangle - E_{\sU\sn\si}\} 
$$
\beq
  = :    
\{\pa_hS\}^2 + A\pa_hS + B \hspace{5.6in}
\label{h-eq} 
\eeq 
(dropping the $\pa_h\mbox{}^2S$ term as per the WKB approximation in the last expression).   
\noindent Applying instead \{1 -- $\mP_{\chi}$\} $\times$ to form the {\it l-equation}, for now of fluctuation equation form \cite{AK13} 
\beq
\{1 - \mP_{\chi}\}  \big\{ - 2i\hbar  \pa_{h} S \, \pa_{h}|\chi\rangle - \hbar^2\big\{  \pa_{h}\mbox{}^2  + 2h^{-1}  \pa_{h} + 
h^{-2}\triangle_{l}\}  |\chi\rangle + 2\{V_{l}(l^{\sfa}) + J(h, l^{\sfa})\}  |\chi\rangle  \big\}   = 0  \mbox{ } .  
\label{l-TDSE-prime}
\eeq
Here, $\mP_{\chi}$ is the projection operator $|\chi\rangle\langle\chi|$.  
Systems of equations of the above general type (\ref{h-eq}, \ref{l-TDSE-prime}) first arose in the work of Banks \cite{Banks} and of Halliwell--Hawking \cite{HallHaw} 
(see Appendix A, though that has discarded a lot of the terms kept above).  

\mbox{ } 

\noindent{\bf Emergent WKB time}. 
It is then standard in the semiclassical approach to use that $\pa_{h}\mbox{}^2 S$ is negligible by the WKB approximation to remove the second term from the $h$-equation, and apply 
\beq
\pa_{h} S = p_{h} = \mbox{\Last}h
\label{lance2}
\eeq
by identifying $S$ as Hamilton's function, and using the expression for momentum in the Hamilton--Jacobi formulation, the momentum-velocity relation, and the chain-rule to 
recast $\pa_{h}$ as $\pa_{h}t\, \Last$. 

\noindent Note 1) This expression is a classical time too, so to this level of approximation, I write simply $t^{\se\sm}$.

\noindent Note 2) \noindent It amounts to recovery of Newtonian time, proper time and cosmic time in suitable contexts.

\noindent Note 3) Only the $h$-change contributes to it: $t^{\se\sm} = F[h, \d h]$.
To improve it to a Machian emergent time, we must give an opportunity for the $l$ to contribute on the timestandard too (`a backreaction').   
Then the $l$-subsystem contributes {\sl differently} in the classical and quantum cases, so their fully Machian timestandards {\sl do not coincide}: 
$t^{\se\sm(\scc\sll)} = F[h, l, \d h, \d l]$ versus $t^{\se\sm(\sW\sK\sB)} = F[h, l, \d h, |\chi(h, l)\rangle]$.  

\noindent Neglecting almost all the terms in (\ref{h-eq}), this $h$-equation collapses to the standard semiclassical approach's Hamilton--Jacobi equation,    
\beq
\{\pa_{h}S\}^2 = 2\{E_h - V_{h}\} \mbox{ } , \mbox{ } \mbox{ } 
\label{Sicut}
{\Last} h^2 = 2\{E_h - V_{h}\} \mbox{ } , \mbox{  } \mbox{ or } \mbox{  }
\left\{
{\Last h}/{h}\right\}^2 = {2E_h}/{h^2} - {2V_{h}}/{h^{2}} \mbox{ }   . 
\eeq
The second form is by (\ref{lance2}), and is especially justified because $S$ is a standard Hamilton--Jacobi function.
The third form is an analogue Friedmann equation.   
Whichever of the above forms is then solved by
\beq
{\lt}^{\se\sm} = 2^{-1/2}  \int  {\d h}/{\sqrt{W_h}}  \mbox{ } , \mbox{ } \mbox{ } W_h := E_h - V_h(h) \mbox{ } .   
\label{t-em-WKB}
\eeq
\noindent{\bf Recasting of the $l$-fluctuation equation as a TDSE} (time-dependent Schr\"{o}dinger equation).
One passes for a fluctuation equation to a semiclassical emergent TDSE via manipulating the crucial chroniferous cross-term, along the following lines \cite{FileR} 
\beq
i\hbar \,  \pa_h S         \,   \pa_h |\chi\rangle  = 
i\hbar \,   p_{h}          \,   \pa_h |\chi\rangle =
i\hbar \frac{  \pa h       }{   \pa t^{\se\sm(\sW\sK\sB)}  }\frac{\pa \left| \chi\right \rangle}{\pa h} \approx  
i\hbar                                                     \frac{\pa \left| \chi\right\rangle }{  \pa t^{\se\sm(\sW\sK\sB)}  }    \mbox{ } , 
\eeq
via (\ref{lance2}) and the chain-rule in reverse.

\mbox{ } 

\noindent{\bf Rectified emergent Machian time}.
Provided that one takes the TDSE core seriously, at least to good approximation, and as opposed to a more general TDWE form, the $l$-equation further simplifies if one chooses the 
{\sl emergent rectified time} \cite{FileR} that is given by
\beq
h^2\Last := h^2 \pa/\pa t^{\se\sm(\sW\sK\sB)} = \pa/\pa t^{\se\sm(\sr\se\scc)} =: \mbox{\textcircled{$\star$}} \mbox{ } .  
\label{Tu}
\eeq
This suggests that, whilst emergent WKB time follows on as a quantum-corrected form of emergent JBB time, 
the mathematics of the quantum system dictates passage to the rectified time instead as regards semiclassical quantum-level calculations.
The zeroth-order approximation to this is 
\beq
\mbox{\large $t$}^{\se\sm(\sr\se\scc)}_{(0)} =  
\int\left.\d t^{\se\sm(\sW\sK\sB)}_{(0)}\right/h^2\big(t^{\se\sm(\sW\sK\sB)}_{(0)}\big) = 2^{-1/2} \int {\d h}/{h^2\sqrt{{W_h}}} 
\mbox{ } . 
\label{Hannah}
\eeq
However, we define (\ref{Tu}) this to arbitrary order, for which it again is a quantum-Machian expression $F[h, l, \d h, |\chi(h, l)\rangle]$, related to the previous such 
by a conformal transformation. 
This amounts to a relationally-motivated freedom \cite{Banal} in choice of timestandard, and so lies within the same theoretical scheme from the Machian perspective).  
Additionally, using $t^{\se\sm(\sr\se\scc)}$ amounts to working on the shape space itself, which is natural from a geometrical perspective.    
Finally, note that GR has a different rectification \cite{AMSS1} that originates from the presence of an $a^3$ factor in the action 
(from a $\sqrt{h}$ factor for $h$ the determinant of the spatial 3-metric).  

\mbox{ } 

\noindent A fairly general form of $l$-TDWE with respect to $t^{\se\sm(\sr\se\scc)}$ is then 
\beq
i\hbar\{1 - \mP_{\chi}\} \Club |\chi\rangle = \{1 - \mP_{\chi}\} \{ - \hbar^2 \{\Spade^2 + \Spade  + \triangle_{l} \}/2 +  V^{\sr\se\scc}_{l} + J^{\sr\se\scc}  \}   |\chi\rangle . 
\label{Rec-TDSE}
\eeq
for $\Club    := \Rec  - \Rec  l\pa_{l}$  and $\Spade := \Club/\Rec  \, \mbox{ln} \, h(t^{\se\sm(\sr\se\scc)})$.  
\noindent Here, one is using (\ref{Hannah}) to express $h$ as an explicit function of $t^{\se\sm(\sr\se\scc)}$ 
This does require invertibility (at least on some intervals of the mechanical motion).
That granted, one can cast the TDWE as a $t^{\se\sm(\sr\se\scc)}$-dependent perturbation equation. 
\noindent (\ref{Rec-TDSE}) is most commonly approximated as a TDSE, possibly treated perturbatively if the $J$ term (or further terms) are kept.   
\beq
i\hbar\mbox{\textcircled{$\star$}}|\chi\rangle = - \hbar^2\triangle_{l}\left|\chi\right\rangle/2 + 
V^{\sr\se\scc}_{l} \left|\chi\right\rangle + {J}^{\sr\se\scc}\left|\chi\right\rangle \mbox{ \{+ further perturbation terms\} } . 
\eeq
\noindent Note 1) For triangleland, this is a mathematically familiar equation. 

\noindent Note 2) A schematic form for this is $\{1 - P_{\chi}\}\{i\hbar\Club - \widehat{H}_l \}|\chi\rangle = 0$ 
and $\widehat{H}_l = \widehat{N}_l - {\hbar^2}\{\Spade^2 + \Spade\}/2$ for 
\beq
\widehat{N}_l := - \hbar^2\triangle_l/2 + V_l^{\sr\se\scc} + J^{\sr\se\scc}
\label{N-def}
\eeq 
the part that neglects any other time derivative terms. 

\mbox{ }

\noindent{\bf Rectified $h$-equation}.
We rectify the $h$-equation too, so as to place the system of equations on a common footing in terms of a single time variable. 
\noindent It is 
$$
0 = \{\Rec \, \mbox{ln} \, h\}^2 - 2i \hbar \{ \langle  \Spade  \rangle  +  1 \} \Rec \, \mbox{ln} \, h  - \hbar^2 \langle \Spade^2 + \Spade  \rangle  
 \hspace{2.5in}
$$
\beq
- \hbar^2\{\langle \triangle_l \rangle - 3/2\} - 2\{E^{\sr\se\scc} -  V_h^{\sr\se\scc} - \langle  V_l^{\sr\se\scc} + J^{\sr\se\scc}  \rangle \} \hspace{0.2in}  = : \hspace{0.2in}
\{\Rec \, \mbox{ln} \, h\}^2 + C \Rec \, \mbox{ln} \, h + D  \mbox{ } .
\label{Rec-h}
\eeq
Its simplest truncation is 
\beq
\{\Rec \, \mbox{ln} \, h\}^2   = 2\{E^{\sr\se\scc} -  V_h^{\sr\se\scc} \} \mbox{ } \mbox{ } , \mbox{ } \mbox{ which integrates to }  \mbox{ } \mbox{ }  
\lt^{\se\sm(\sr\se\scc)} = \int \d h\left/h\sqrt{2W^{\sr\se\scc}}\right. = \int {\d h}\left/{h^2\sqrt{2W_h}}\right. \mbox{ } .  
\label{EE-2}
\eeq
See \cite{AMSS1, AMSS2} for GR minisuperspace counterparts of these final-form semiclassical quantum cosmology equations.

\section{Brief term-by-term outline, with names of regimes looked at so far}

\noindent {\bf Adiabatic terms}. We already discussed these in Sec 3.  

\noindent{\bf Backreaction terms} These concern the $l$-subsystem induced corrections to the $h$-subsystem \cite{Kieferbook}. 
These are motivated by 1) book-keeping: without them, the model for the $l$-subsystem can lose energy without the $h$-subsystem gaining an equal amount.  
2) Back-reaction is conceptually central to GR.  
3) By wishing to realize the Machian opportunity for $l$-change to contribute to the timestandard for the whole universe.  

\noindent {\bf Higher derivative terms}. 
One often neglects the extra $t^{\se\sm(\sW\sK\sB)}$-derivative terms whether by discarding them prior to noticing they are also convertible 
into $t^{\se\sm(\sW\sK\sB)}_{(0)}$-derivatives or by arguing that $\hbar^2$ is small or $\rho$ variation is slow. 
But one would expect there to be some regions of configuration space where the emergent TDWE behaves more like a Klein--Gordon equation than a TDSE, albeit in full it is more general.
Moreover there is a potential danger in ignoring higher derivative terms even if they are small (c.f. Navier--Stokes equation versus Euler equation in fluid dynamics).
The aforementioned $O((H/m_{\sP\sll})^2)$ effect can be seen to arise from such considerations in parallel \cite{KS91} to the {\sl second} approximation to relativistic QM's 
Klein--Gordon equation (the TDSE being the first). 
It arises from making the substitution $\pa_t^2 \approx \widehat{H}_l^2$.  

\noindent {\bf Averaged terms}. These are the subject of the next Sec. 
Expectation/averaged terms are often dropped in the Quantum Cosmology literature,\footnote{Incidentally, the idea of neglecting averaged terms illustrates how dimensional analysis is 
not all, since averaged and unaveraged versions of a quantity clearly have the same `dimensionless groups' (in the sense used in Fluid Mechanics).}  
%
via these are argued to be negligible by the {\it Riemann--Lebesgue Theorem}: the mathematics corresponding to the physical idea of {\it destructive interference}.  
 
\mbox{ }

\noindent \bu In the simplest regime, one drops all the correction terms, so one is left with, firstly, the Hamilton--Jacobi equation (\ref{Sicut}) to solve 
for one's emergent time function, and, secondly, a TDSE with respect to this time.

\noindent \bu In the second-simplest regime, one does keep an interaction term on the $l$-equation, and treats this as a $t$-dependent perturbation of that TDSE.
This was solved for 3-particle RPM's in 1- and 2-$d$ in \cite{SemiclI, SemiclIII}.  

\noindent \bu However, one then has the objections that I motivated back-reaction with.  
Thus, one should include a means of back-reaction. 
In the simplest such case, let that be small too.  
One now has a coupled perturbative scheme.
Moreover, it is a hybrid of classical and quantum perturbation theory. 
And stranger still if the time is to be determined by it, for, until determined, time is the most dependent variable of them all and therefore also subject to perturbation.  
See \cite{ACos2} for details of the equations of this simplest perturbative realization of Machian emergent semiclassical time.  

\noindent \bu The more significant types of contribution to the Machian semiclassical time are as follows (keeping up to 1 power of $\hbar$) 
\beq
\lt^{\se\sm(\sr\se\scc)} = \lt^{\se\sm(\sr\se\scc)}_{(0)} + \frac{1}{2\sqrt{2}}\int\frac{\langle  J  \rangle}{W_{h}^{3/2}}\frac{\d h}{h^2} - 
\frac{i\hbar}{2}  \int   \frac{\d h}{h^2 W_{h}}  \left\{  \frac{1}{h} + \left\langle \frac{\pa}{\pa h} \right\rangle \right\} + O(\hbar^2) \mbox{ } .   
\label{QM-expansion}
\eeq
One can see that these corrections are quantum in origin: an operator-ordering term and two expectation terms.  
Comparing with the classical expression for $l$-corrections to the $h$-time, 
\beq
\lt^{\se\sm(\sJ\sB\sB)}_{(1)} = \lt^{\se\sm(\sJ\sB\sB)}_{(0)} + \frac{1}{2\sqrt{2}}\int \frac{\d h}{\sqrt{W_{h}}}
\left\{
\frac{J}{W_h} + 
\left\{  
\frac{    ||\d \bl||_{\sbM}   }{  \d\{ \mbox{ln} \, h \}   }
\right\}^2
\right\} 
+ O\left(\left(  \frac{J}{W_h}   \right)^2\right) 
+ O\left(\left(  \frac{||\d \bl||_{\sbM}}{\d\{ \mbox{ln} \, h\}}\right)^4\right) 
\label{Cl-Expansion}
\eeq
for shape variables $l^{\sfa} = S^{\sfa}$ and shape part of the configuration space metric $\bM$.
Here, one can see that interaction has been replaced by expectation of interaction and two quantum terms with no classical counterpart have arisen alongside it.  
The differences by factors of $h^2$ in the integrands are of course due to the rectification, which one can view as an affine transformation. 
See \cite{FileR, ACos2, AK13} for computation of these various classical and quantum terms by knowing the potentials and the wavefunctions and evaluating the integrals. 
See \cite{AMSS1, AMSS2} for corresponding minisuperspace calculations.     
Note that $O((\mbox{shape}/\mbox{scale})^2)$ effects ensue in the above working (whether in terms of RPM shape or of anisotropy or inhomogenity in the GR setting).  

\noindent \bu Finally, note that, whilst adiabatic and higher-derivative corrections have been considered as per above for a few minisuperspace examples, 
they have not yet been done for RPM's. 
On the other hand, a satisfactory solving-treatment of non-negligible averages has not to our knowledge been attempted yet.  
The next Sec brings us one step closer to attaining this.

\section{Motivation for the averaged terms}\label{Av}

\subsection{They are not always small}\label{Issue}

There are, however, some reasons to keep these on some occasions.

\noindent 1) A triangleland counterexample to such terms being small is as follows. 
For the free problem, $\langle D^2\rangle|\chi\rangle$ and $D^2|\chi\rangle$ are of the same size since the wavefunctions in question are eigenfunctions of this operator.

\noindent 2) Such terms turned out to be important in Atomic and Molecular Physics, as per the next SSec.

\subsection{Hartree--Fock approach is crucial in Atomic and Molecular Physics}

\noindent Here \cite{HF, AtFFF}, one considers trial wavefunctions of the form $\psi^0 = \mbox{det}\big(\phi_{a_1}(1) ... \phi_{a_n}(n)\big)/\sqrt{n}$ 
for $\phi_{a_I}(I)$ the wavefunctions of decoupled atoms.
The decoupled Hamiltonian $H^0 = \sum_{I=1}^n h_I$ for individual decoupled atom Hamiltonians $h_I := -\{\hbar^2/2m_e\}\nabla^2 + V_{\mbox{\scriptsize fixed nuclear background}}$.  
(We can see this is compatible with the semiclassical approach insofar as a BO approximation is in use in both)
We next wish to include electron--electron repulsion terms $H^{ee} := \half \sum\sum_{I \neq J} e^2/ 4 \pi \epsilon_0 r_{IJ}$.  
Then $\langle \psi | H^0 | \psi \rangle = n\langle \psi | h_1 | \psi \rangle$ since the electrons are indistinguishable, and so
\beq
\langle \psi | \sum_{\mbox{ } \, \, I}\sum_{\neq \, \, \, J}e^2/4\pi\epsilon_0 r_{IJ} | \psi \rangle = \half \, n\{n - 1\} \langle \psi | e^2/4\pi\epsilon_0r_{12} | \psi \rangle = 
\half \sum_{\mbox{ } \, \, I}\sum_{\neq \, \,\, J} [\phi_I\phi_J|\phi_I\phi_J]
\eeq
for $[\phi_I\phi_J|\phi_I\phi_J] := \langle \phi_I^{(1)}\phi_J^{(2)} | e^2/4\pi\epsilon_0r_{IJ} | \phi_I^{(1)}\phi_J^{(2)} - \phi_J^{(1)}\phi_I^{(2)} \rangle$. 
Then, proceeding variationally, minimize 
\beq
E = \sum_{I = 1}^n \langle \phi_I | h_1 | \phi_J \rangle + \half \sum_I\sum_J\{ [\phi_I\phi_J|\phi_I\phi_J] - [\phi_I\phi_I|\phi_J\phi_J]  \}
\eeq
subject to the orthogonality of the decoupled wavefunctions,
\beq
\sum_I\sum_J \langle \phi_I | \phi_J \rangle - \delta_{IJ} = 0 
\eeq
(incorporated by the method of Lagrange multipliers).
This gives, as equation to solve, 
\beq
h_1\phi_I(1) + \sum_{J = 1}^N \hat{J}_J(1)\phi_I(1) - \hat{K}_J(1)\phi_I(1)\} = \sum_{J = 1}^N \lambda_{JI}\phi_J(1) \mbox{ } , 
\eeq
for $\hat{J}_J(1) := \langle \phi_J(2) | e^2/4\pi\epsilon_0 r_{12}  | \phi_J(2) \rangle$ the {\it Coulomb operator} 
and $\hat{K}_J$ defined by $K_J(1)\phi_I(1) := \langle \phi_J(2) | e^2/4\pi\epsilon_0 r_{12}  | \phi_I(2) \rangle \phi_J(1)$ the {\it exchange operator}.
The main points to be made are that 1) these are coupled integro-differential equations (just like semiclassical quantum cosmology backreactions), 
and these are known to require a numerical treatment. 
Namely, a `self-consistent method': a type of iterative procedure consisting of repeatedly cycling around substituting approximate solutions to equations into other equations.
2) {\sl Not} taking these averaged terms into account does produce a simpler set of equations to solve, but is substantially inaccurate as regards atomic and molecular spectra.

\section{Variational principles for quantum cosmology}

While there are a number of differences between Molecular Physics and Quantum Cosmology, Hartree--Fock theory and the variational principles upon which it is based 
in fact is known to span those differences.

\noindent A) It is available in situations that make use of a simple- rather than antisymmetrized-product wavefunction; 
in fact the first of each of Hartree and Slater's cited works used the simple product, Slater being the provider of the underlying variational principle.  
This corresponds to a non-fermionic ansatz, which is in fact the status quo for Quantum Cosmology, if not of course in Atomic and Molecular Physics.

\subsection{Variational principles for time-independent QM} 

These are well-known.  
There are two types: 1) the Ritz Principle
\beq
E[\psi, \psi^*] = \langle \psi | H | \psi \rangle/\langle \psi | \psi \rangle \mbox{ } ,
\eeq
for which varying with respect to $\psi^*$ returns the TISE, $H|\psi\rangle = E |\Psi\rangle$.
2) An alternative variational principle for attaining this is
\beq
J[\psi, \psi^*] = \langle \psi| H |\psi\rangle - E ||\psi||^2 \mbox{ } , 
\eeq
where $E$ is a Lagrange multiplier that encodes the normalization condition $||\psi||^2 = 1$.

\noindent B) There is no problem in extending such principles to curved (and, for GR, indefinite) configuration spaces.  
Then the specific principle of type 2) for the reduced formulation of triangleland is 
\beq
\int_{\sfR(3, 2)}{\cal D}\fR(3, 2)
\{\Psi^* \{-\hbar^2\{\pa^2_{\rho} + 2\rho^{-1}\pa_{\rho} + \rho^{-2} \{ \triangle_{\mathbb{CP}^1} - 3/2\}\} + V(\rho, \Theta, \Phi)\}\Psi + E_{\sU\sn\si}|\Psi|^2\} \mbox{ } .
\eeq
On the other hand, for GR minisuperspace with a scalar field, it is (using $\alpha := \mbox{ln}\,a$)
\beq
\int_{\mbox{\scriptsize MSS}}{\cal D} (\mbox{MSS}) 
\left\{ 
-\hbar^2\Psi^*\{\pa_{\alpha}^2 - \pa_{\phi}^2\}\Psi + \mbox{e}^{6\alpha}m^2\phi^2 - \mbox{e}^{4\alpha}\}|\Psi|^2
\right\}
\mbox{ } 
\eeq 

\subsection{Variational principles for time-dependent QM} 

\noindent C) Moreover, $t$-dependent quantum variational principles are also known.
Here, the successors of the Ritz and multiplier principles are distinct \cite{BR86}, and it is the latter that is more suitable to our purposes, namely 
\beq
S[\psi,\psi^*] = \int \d V \psi^* \{ i\hbar\pa_t - H - \Lambda\}\psi \mbox{ } .  
\label{TDSE-VP}
\eeq
$\Lambda$ is associated both with normalization and with freedom to change phase factor.
Time-dependent Hartree--Fock theory has been considered in e.g. \cite{Dirac30, KK76, BR86}.

\subsection{Variational principle for $l$-TDSE of semiclassical quantum cosmology} 

\noindent We next apply this to encode the semiclassical $l$-equation by itself:
\beq
S[\psi,\psi^*] = \int \d V \chi^* \{       i\hbar\Club - \widehat{N}_l - \half\{i\hbar\langle \Club \rangle - \langle \widehat{N}_l \rangle \} - 
         \{\Lambda - \langle\Lambda\rangle\}  \}  \chi        
\label{l-VP}
\eeq
using for now the $t$-derivative-free $\widehat{N}_l$ as an approximand for $\widehat{H}_l$.  
Now, phases cancel, but $\Lambda$ still variationally encodes the normalization condition.  
Post-variationally, however, one can recognise that the $\Lambda - \langle\Lambda\rangle$ combination plays the role of a zero multiplier.  
This scheme assumes a satisfactory $t^{\se\sm(\sr\se\scc)}$ has already been determined. 
Then variation with respect to $\chi^*$ here encodes eq (\ref{Rec-TDSE}).   
Read off eq (\ref{N-def}) to have the triangleland case; for minisuperspace counterparts of this, see \cite{AMSS2}. 

\noindent D) Next, note that variational principles similar to (\ref{TDSE-VP} exist for other TDWE's, 
so, ultimately, including $-\hbar^2\{\Spade^2 + \Spade\}/2$ corrections into the variational scheme is not a problem either. 
Just add $-\hbar^2\{\{\Spade^2 + \Spade\}/2 - \langle\Spade^2 + \Spade\rangle/4\}$ into the innermost factor of (\ref{l-VP}).

\subsection{Further modelling features}

\noindent E) In anticipation of applying such methods to Halliwell--Hawking type semiclassical quantum equations themselves, 
I mention that the Hartree--Fock method has been set up for field theories (see e.g. \cite{KK76, Field-HF}).  

\noindent F) For full GR (or the Dirac presentation of 2-$d$ scaled RPM's), one needs to deal with linear constraints, the treatment of which I difer to \cite{SemiclV}.

\subsection{Variational principles for $h$-equation of semiclassical quantum cosmology}

\noindent For $h$ alone, can use some corrected form of classical action principle, of the schematic form
\beq
S = \int\d t^{\se\sm(\sr\se\scc)}\{T - V - \langle \widehat{O} \rangle\}
\eeq
(see \cite{SemiclV} for details).
However to couple $h$ and $l$ together, it is more convenient to encode the solution $S = S[t^{\se\sm(\sr\se\scc)}, l, |\chi(t^{\se\sm(\sr\se\scc)}, l\rangle]$ 
rather than the Hamilton--Jacobi equation itself directly. 
Then the $h$-only variational principle is 
\beq
S =  \int \d t^{\se\sm(\sr\se\scc)}            \{           -A \pm \sqrt{A^2 - B}   \}^2/h^2 \mbox{ } .
\eeq
This goes more like the conventional `Hamilton function = action' in the case with negligible linear terms $A$.

\subsection{Variational principles for $h$-$l$ coupled system of semiclassical quantum cosmology}

Schematically, the coupled $h$--$l$ system is of the form 
\beq
\mbox{\Huge\{}  \stackrel{\mbox{( chroniferous Hamilton--Jacobi $h$-equation with expectation corrections)}}
                         {\mbox{(emergent-time-dependent Hartree--Fock $l$-equation)}}
                          \mbox{ } , 
\label{cyril}  
\eeq
which is probably this time a new type of system from a Mathematical Physics perspective.  
A {\sl non-chroniferous} variational principle for this system is
$$
S[\psi,\psi^*] = \int \d t^{\se\sm(\sr\se\scc)} 
\int \d V \chi^* \{       i\hbar       \Club        - \hbar^2\{       \Spade^2            +        \Spade        \} - \widehat{N}_l - 
                   \half\{i\hbar\langle\Club\rangle - \hbar^2 \langle \Spade^2            +        \Spade \rangle   - \widehat{N}_l \rangle \} - 
\{\Lambda    -    \langle\Lambda\rangle\}   \}   \chi
$$
\beq
+ \bigmu 
\left\{
S -  \int \d t^{\se\sm(\sr\se\scc)} \{-A \pm \sqrt{A^2 - B}\}^2/h^2
\right\} 
\mbox{ } . 
\eeq
Note that the $\langle \mbox{ } \rangle$ in the expression encoding the $h$-equation does not enter the $l$-equation since variation with respect to $S$ 
yields the post-variational condition on the multiplier $\bigmu = 0$.
Thus indeed this encodes eqs (\ref{Rec-h}) and (\ref{Rec-TDSE}).  

\noindent Note also that this is indeed still assuming we know $t$; if not, one could formulate a variational principle that gives equations that do not refer to $t$.  
Then one only has $t$ as an expression to judge by, iteration by iteration, until a satisfactory accuracy is attained.
That is somewhat lengthy and requires explaining relational formalism in more detail, so I leave its exposition for another occasion.

\section{Conclusion}\label{Conclusion}

Quantum Cosmology equations contain many terms.
One of these produces $O((H/m_{\sp\sll})^2)$ effect, much larger than other quantum-gravitational effects robustly predicted to date.
But many other terms in the semiclassical equations are neglected rather than investigated regime by regime.
This is in part because, there being many terms, there are very many regimes that keep combinations of these terms.
Thus I consider qualitative toy models with simple maths, for which I have considered some of the regimes.

\noindent Once one needs to work approximately in QM, two of the first ports of call are perturbation methods and variational methods. 
In semiclassical Quantum Cosmology, there is a nontrivial merging of classical and quantum concepts for perturbations and variations, 
which is considered for perturbations in \cite{ACos2, AMSS2} and for variational methods in the present article and \cite{SemiclV}.

\noindent What is to be used as the variational trial wavefunction in Quantum Cosmology?  
A plain product ansatz is the simplest, but a latter-day Slater might suggest a better from physical principles concerning the nature of inhomogeneous quantum 
fluctuations in the early universe.
\noindent For the Halliwell--Hawking model, one can split the wavefunction up mode by mode (\ref{MBM}) and the wavefunction for each mode is itself a product of 
scalar (S), vector (V), tensor (T) parts (\ref{SVT}).
This is some sort of variational trial wavefunction parallelling the plain (as opposed to Slater antisymmetric fermionic) Hartree--Fock trial wavefunction.
In this regard, triangleland falls rather short as a model for a new fourth reason after those listed in the Introduction: it has no mode expansion or S--V--T decomposition. 
One can have a smaller product-type trial wavefunction, $\psi_{\sp\se\sr\st} = \psi(\delta\rho)\psi(\delta\Theta)\psi(\delta\Phi)$. 
Including S--V--T couplings in the Halliwell--Hawking model might bear some parallel with including $e^{-}$--$e^-$ terms in the Hartree--Fock approach to Atomic/Molecular Physics.

\noindent Finally, one would need to face whether time-dependent Hartree--Fock type schemes for Quantum Cosmology iteratively converge. 
%

\mbox{ } 

\noindent {\bf Acknowledgements}: 
I thank the 'XXIX-th International Workshop on High Energy Physics: New Results and Actual Problems in Particle \& Astroparticle Physics and Cosmology' (Moscow 2013) 
for invitation to Speak and to run the Cosmology Discussion session.   
I thank Marc Lachieze-Rey, Claus Kiefer, Jeremy Butterfield, Reza Tavakol, Malcolm MacCallum and Don Page for help with my career.
This work was started whilst a Research Fellow at Peterhouse in 2007.  
I developed this work further in 2012 under the support of a grant from the Foundational Questions Institute (FQXi) Fund, 
a donor-advised fund of the Silicon Valley Community Foundation on the basis of proposal FQXi-RFP3-1101 to the FQXi.  
I thank also Theiss Research and the CNRS for administering this grant.  

\begin{appendices}

\section{Halliwell--Hawking (HH) model}\label{HH}

Perform an ADM split of GR's metric and its action (including a minimally-coupled scalar field $\phi$).  
Consider approximately homogeneous isotropic cosmologies in the sense of \cite{HallHaw}
\beq
h_{ij} = a(t)^2\{\Omega_{ij} + \epsilon_{ij}\} \mbox{ } , \mbox{ } \phi = \sigma^{-1}\big\{\phi(t) + \sum_nf_nQ^n \big\} \mbox{ } . 
\eeq
Here, $h_{ij}$ is the spatial 3-metric, $a$ is the approximate cosmological scalefactor, $\Omega_{ij}$ is the $\mathbb{S}^3$ metric 
(so it is a closed cosmology and anisotropy is being ignored) and $\epsilon_{ij}$ are inhomogeneous perturbations of the schematic form
\beq
\epsilon_{ij} = \sum_n\big\{ a_n\Omega_{ij}Q^n + b_nP_{ij}^n + c_nS_{ij}^n + d_nG_{ij}^n \big\} \mbox{ } , 
\eeq
for $Q^n$, $P^n$, $S^n$ and $G^n$ various kinds of tensor harmonics (with indices suppressed).
Also, $\sigma := \sqrt{2/3\pi}/m_{\sP\sll}$ (normalization factor);  the coefficients $a_n$, $b_n$, $c_n$, $d_n$, $f_n$ are functions of $t$ alone.
Then the Semiclassical Approach yields, after a number of approximations of the qualitative kinds discussed in the main text, separated-out TISE's 
(time-independent Schr\"{o}dinger equation's) mode by mode, 
\beq
i\hbar\frac{\pa\Psi^{(n)}}{\pa t^{\se\sm(\sW\sK\sB)}} = \widehat{H}^{(n)}_{\mbox{\scriptsize 2nd order}}\Psi^{(n)} \mbox{ } . 
\label{MBM}
\eeq
Furthermore, 
\beq
\Psi^{(n)} = \Psi^{(n)}_{\mbox{\scriptsize scalar}} \Psi^{(n)}_{\mbox{\scriptsize vector}} \Psi^{(n)}_{\mbox{\scriptsize tensor}} \mbox{ } .
\label{SVT}
\eeq
$$
\widehat{H}^{(n)}_{\mbox{\scriptsize 2nd order}} := \frac{e^{-3\alpha}}{3}
\left\{
\left\{
\frac{a_n^2}{2} + \frac{10\{n^2 - 4\}}{n^2 - 1}b_n^2
\right\}
\pi_{\alpha}^2 +
\left\{
\frac{15a_n^2}{2} + \frac{6\{n^2 - 4\}}{n^2 - 1}b_n^2
\right\}
\pi_{\phi}^2
- \pi_{a_n}^2 + \frac{n^2 - 1}{n^2 - 4}\pi_{b_n}^2 + \pi_{f_n}^2 
\right.
$$
$$
+ 2a_n\pi_{a_n}\pi_{\alpha} + 8b_n\pi_{b_n}\pi_{\alpha} - 6a_n\pi_{f_n}\pi_{\phi} 
- e^{4\alpha}
\left\{
\frac{n^2 - 5/2}{3}a_n^2 + \frac{\{n^2 - 7\}\{n^2 - 4\}}{3\{n^2 - 1\}}b_n^2 + \frac{2\{n^2 - 4\}}{3}a_nb_n - \{n^2 - 1\}f_n^2 
\right\}
$$
\beq
\left. 
+ e^{6\alpha}m^2\{f_n^2 + 6a_nf_n\phi\} + e^{6\alpha}m^2\phi^2
\left\{
\frac{3a_n^2}{2} - \frac{6\{n^2 - 4\}}{\{n^2 - 1\}}b_n^2
\right\}
\right\}
\eeq
for $\pi_q$ the momentum conjugate to whatever quantity $q$.

\end{appendices}  


\end{document}